\documentclass[12pt,a4paper]{article}

\usepackage{float}
\makeatletter
\let\@tmp\@xfloat     
\usepackage{fixltx2e}
\let\@xfloat\@tmp                    
\makeatother

\usepackage{ifthen} 
\usepackage{longtable} 
\newboolean{pdflatex}
\usepackage{xcolor}
\usepackage{rotating}
\setboolean{pdflatex}{true} 

\newboolean{articletitles}
\setboolean{articletitles}{true} 

\newboolean{uprightparticles}
\setboolean{uprightparticles}{false} 

\newboolean{inbibliography}
\setboolean{inbibliography}{false} 

\def\paperauthors{LHCb collaboration} 
\def\paperasciititle{Search for the rare decay $\Lambda_{c}^{+} \to p\mu^+\mu^-$}
\def\papertitle{Search for the rare decay $\Lc \to p\mu^+\mu^-$}
\def\paperkeywords{{High Energy Physics}, {LHCb}} 
\def\papercopyright{CERN on behalf of the LHCb collaboration}
\def\paperlicence{CC-BY-4.0}
\def\paperlicenceurl{https://creativecommons.org/licenses/by/4.0/}

\usepackage[top=1in, bottom=1.25in, left=1in, right=1in]{geometry}

\usepackage{microtype}
\usepackage{lineno}  
\usepackage{xspace} 
\usepackage{caption} 

\usepackage{float}

\usepackage{graphicx}  
\usepackage{color}
\usepackage{colortbl}
\graphicspath{{./figs/}} 

\usepackage{amsmath} 
\usepackage{amssymb}
\usepackage{amsfonts}
\usepackage{upgreek} 

\newcommand*\patchAmsMathEnvironmentForLineno[1]{%
\expandafter\let\csname old#1\expandafter\endcsname\csname #1\endcsname
\expandafter\let\csname oldend#1\expandafter\endcsname\csname
end#1\endcsname
 \renewenvironment{#1}%
   {\linenomath\csname old#1\endcsname}%
   {\csname oldend#1\endcsname\endlinenomath}%
}
\newcommand*\patchBothAmsMathEnvironmentsForLineno[1]{%
  \patchAmsMathEnvironmentForLineno{#1}%
  \patchAmsMathEnvironmentForLineno{#1*}%
}
\AtBeginDocument{%
\patchBothAmsMathEnvironmentsForLineno{equation}%
\patchBothAmsMathEnvironmentsForLineno{align}%
\patchBothAmsMathEnvironmentsForLineno{flalign}%
\patchBothAmsMathEnvironmentsForLineno{alignat}%
\patchBothAmsMathEnvironmentsForLineno{gather}%
\patchBothAmsMathEnvironmentsForLineno{multline}%
\patchBothAmsMathEnvironmentsForLineno{eqnarray}%
}

\usepackage{hyperxmp}

\usepackage[pdftex,
            pdfauthor={\paperauthors},
            pdftitle={\paperasciititle},
            pdfkeywords={\paperkeywords},
            pdfcopyright={Copyright (C) \papercopyright},
            pdflicenseurl={\paperlicenceurl}]{hyperref}

\usepackage[all]{hypcap} 

\usepackage{xspace} 
\usepackage{upgreek}

\def\lhcb {\mbox{LHCb}\xspace}

\def\MagUp {\mbox{\em Mag\kern -0.05em Up}\xspace}

\ifthenelse{\boolean{uprightparticles}}%
{

 \def\Pmu         {\ensuremath{\upmu}\xspace}

 \def\Ppi         {\ensuremath{\uppi}\xspace}

 \def\Pphi        {\ensuremath{\upphi}\xspace}

 \def\Pomega      {\ensuremath{\upomega}\xspace}                 

 \def\PDelta      {\ensuremath{\Delta}\xspace}                 
 \def\PXi      {\ensuremath{\Xi}\xspace}                 
 \def\PLambda      {\ensuremath{\Lambda}\xspace}                 
 \def\PSigma      {\ensuremath{\Sigma}\xspace}                 
 \def\POmega      {\ensuremath{\Omega}\xspace}                 
 \def\PUpsilon      {\ensuremath{\Upsilon}\xspace}                 
 
 \def\PB      {\ensuremath{\mathrm{B}}\xspace}                 
                  
 \def\PD      {\ensuremath{\mathrm{D}}\xspace}

 \def\PK      {\ensuremath{\mathrm{K}}\xspace}

 \def\Pb      {\ensuremath{\mathrm{b}}\xspace}                 
 \def\Pc      {\ensuremath{\mathrm{c}}\xspace}

 \def\Pi      {\ensuremath{\mathrm{i}}\xspace}

 \def\Pp      {\ensuremath{\mathrm{p}}\xspace}

 \def\Ps      {\ensuremath{\mathrm{s}}\xspace}

}
{

 \def\Pmu         {\ensuremath{\mu}\xspace}

 \def\Ppi         {\ensuremath{\pi}\xspace}

 \def\Pphi        {\ensuremath{\phi}\xspace}

 \def\Pomega      {\ensuremath{\omega}\xspace}                 
 \mathchardef\PDelta="7101
 \mathchardef\PXi="7104
 \mathchardef\PLambda="7103
 \mathchardef\PSigma="7106
 \mathchardef\POmega="710A
 \mathchardef\PUpsilon="7107
                  
 \def\PB      {\ensuremath{B}\xspace}                 
                  
 \def\PD      {\ensuremath{D}\xspace}

 \def\PK      {\ensuremath{K}\xspace}

 \def\Pb      {\ensuremath{b}\xspace}                 
 \def\Pc      {\ensuremath{c}\xspace}

 \def\Pi      {\ensuremath{i}\xspace}

 \def\Pp      {\ensuremath{p}\xspace}

 \def\Ps      {\ensuremath{s}\xspace}

}

\makeatletter
\ifcase \@ptsize \relax
  \newcommand{\miniscule}{\@setfontsize\miniscule{4}{5}}
\or
  \newcommand{\miniscule}{\@setfontsize\miniscule{5}{6}}
\or
  \newcommand{\miniscule}{\@setfontsize\miniscule{5}{6}}
\fi
\makeatother

\DeclareRobustCommand{\optbar}[1]{\shortstack{{\miniscule (\rule[.5ex]{1.25em}{.18mm})}
  \\ [-.7ex] $#1$}}

\def\mup        {{\ensuremath{\Pmu^+}}\xspace}
\def\mun        {{\ensuremath{\Pmu^-}}\xspace} 
\def\mumu       {{\ensuremath{\Pmu^+\Pmu^-}}\xspace}

\def\ellell     {\ensuremath{\ell^+ \ell^-}\xspace}

\def\squark    {{\ensuremath{\Ps}}\xspace}

\def\cquark    {{\ensuremath{\Pc}}\xspace}

\def\bquark    {{\ensuremath{\Pb}}\xspace}

\def\pion   {{\ensuremath{\Ppi}}\xspace}

\def\pim    {{\ensuremath{\pion^-}}\xspace}

\def\kaon    {{\ensuremath{\PK}}\xspace}
  \def\Kbar    {{\kern 0.2em\overline{\kern -0.2em \PK}{}}\xspace}

\def\KorKbar    {\kern 0.18em\optbar{\kern -0.18em K}{}\xspace}

\def\Kp      {{\ensuremath{\kaon^+}}\xspace}

\def\Kstarz  {{\ensuremath{\kaon^{*0}}}\xspace}

\def\Kstar   {{\ensuremath{\kaon^*}}\xspace}

  \def\Dbar    {{\kern 0.2em\overline{\kern -0.2em \PD}{}}\xspace}
\def\D       {{\ensuremath{\PD}}\xspace}

\def\DorDbar    {\kern 0.18em\optbar{\kern -0.18em D}{}\xspace}

\def\Dp      {{\ensuremath{\D^+}}\xspace}

\def\Ds      {{\ensuremath{\D^+_\squark}}\xspace}

\def\B       {{\ensuremath{\PB}}\xspace}
\def\Bbar    {{\ensuremath{\kern 0.18em\overline{\kern -0.18em \PB}{}}}\xspace}

\def\BorBbar    {\kern 0.18em\optbar{\kern -0.18em B}{}\xspace}
\def\Bz      {{\ensuremath{\B^0}}\xspace}

  \def\Y#1S{\ensuremath{\PUpsilon{(#1S)}}\xspace}

\def\proton      {{\ensuremath{\Pp}}\xspace}

\def\Lz          {{\ensuremath{\PLambda}}\xspace}
\def\Lbar        {{\ensuremath{\kern 0.1em\overline{\kern -0.1em\PLambda}}}\xspace}
\def\LorLbar    {\kern 0.18em\optbar{\kern -0.18em \PLambda}{}\xspace}

\def\Lc      {{\ensuremath{\Lz^+_\cquark}}\xspace}

\newcommand{\decay}[2]{\ensuremath{#1\!\to #2}\xspace}         

\def\to                 {\ensuremath{\rightarrow}\xspace}

\def\AT#1     {\ensuremath{A_{\mathrm{T}}^{#1}}\xspace}           

\def\C#1      {\ensuremath{\mathcal{C}_{#1}}\xspace}                       
\def\Cp#1     {\ensuremath{\mathcal{C}_{#1}^{'}}\xspace}                    
\def\Ceff#1   {\ensuremath{\mathcal{C}_{#1}^{\mathrm{(eff)}}}\xspace}        
\def\Cpeff#1  {\ensuremath{\mathcal{C}_{#1}^{'\mathrm{(eff)}}}\xspace}       
\def\Ope#1    {\ensuremath{\mathcal{O}_{#1}}\xspace}                       
\def\Opep#1   {\ensuremath{\mathcal{O}_{#1}^{'}}\xspace}                    

\newcommand{\tev}{\ifthenelse{\boolean{inbibliography}}{\ensuremath{~T\kern -0.05em eV}}{\ensuremath{\mathrm{\,Te\kern -0.1em V}}}\xspace}
\newcommand{\gev}{\ensuremath{\mathrm{\,Ge\kern -0.1em V}}\xspace}
\newcommand{\mev}{\ensuremath{\mathrm{\,Me\kern -0.1em V}}\xspace}
\newcommand{\kev}{\ensuremath{\mathrm{\,ke\kern -0.1em V}}\xspace}
\newcommand{\ev}{\ensuremath{\mathrm{\,e\kern -0.1em V}}\xspace}
\newcommand{\gevc}{\ensuremath{{\mathrm{\,Ge\kern -0.1em V\!/}c}}\xspace}
\newcommand{\mevc}{\ensuremath{{\mathrm{\,Me\kern -0.1em V\!/}c}}\xspace}
\newcommand{\gevcc}{\ensuremath{{\mathrm{\,Ge\kern -0.1em V\!/}c^2}}\xspace}
\newcommand{\gevgevcccc}{\ensuremath{{\mathrm{\,Ge\kern -0.1em V^2\!/}c^4}}\xspace}
\newcommand{\mevcc}{\ensuremath{{\mathrm{\,Me\kern -0.1em V\!/}c^2}}\xspace}

\def\mum  {\ensuremath{{\,\upmu\mathrm{m}}}\xspace}

\def\invfb   {\ensuremath{\mbox{\,fb}^{-1}}\xspace}

\def\gsim{{~\raise.15em\hbox{$>$}\kern-.85em
          \lower.35em\hbox{$\sim$}~}\xspace}
\def\lsim{{~\raise.15em\hbox{$<$}\kern-.85em
          \lower.35em\hbox{$\sim$}~}\xspace}

\def\pt         {\mbox{$p_{\mathrm{ T}}$}\xspace}

\def\evtgen     {\mbox{\textsc{EvtGen}}\xspace}

\def\geant      {\mbox{\textsc{Geant4}}\xspace}

\def\photos     {\mbox{\textsc{Photos}}\xspace}

\def\pythia     {\mbox{\textsc{Pythia}}\xspace}

\def\tell1  {TELL1\xspace}
\def\ukl1   {UKL1\xspace}

\newcommand{\ie}{\mbox{\itshape i.e.}\xspace}

\usepackage{cite} 
\usepackage{mciteplus}

\def\LcToPmm {\decay{\Lc}{\proton\mup\mun}}

\def\LcTopmumu  {\decay{\Lc}{\proton \mumu}}
\def\LcTopphi  {\decay{\Lc}{\proton \Pphi (\mumu)}}
\def\LcTopphiNOMU  {\decay{\Lc}{\proton \Pphi(1020) }}

\def\LcTopomegaNM  {\decay{\Lc}{\proton \Pomega }}
\newcommand{\BRof}[1]{\ensuremath{{\cal B}(#1)}\xspace}

\def\Lcpphino   {\decay{\Lc}{\proton \Pphi }}                                                                                                     
                                                                                                 
\def\phimumu  {\decay{\phi}{\mumu}}                                                                                                                           
\def\omegamumu  {\decay{\omega}{\mumu}}                                                                                                                       
\def\Lcpphino  {\decay{\Lc}{\proton \Pphi}}                                                                                                             
\def\Lcpomegano  {\decay{\Lc}{\proton \Pomega}}           
\def\LcpV {\decay{\Lc}{\proton V}} 
\def\LcpVmm {\decay{\Lc}{\proton V (\mup\mun)}} 
\def\CLs {{CL$_{\rm s}$}\xspace}

\usepackage{longtable} 

\begin{document}

\renewcommand{\thefootnote}{\fnsymbol{footnote}}
\setcounter{footnote}{1}

\begin{titlepage}
\pagenumbering{roman}

\vspace*{-1.5cm}
\centerline{\large EUROPEAN ORGANIZATION FOR NUCLEAR RESEARCH (CERN)}
\vspace*{1.5cm}
\noindent
\begin{tabular*}{\linewidth}{lc@{\extracolsep{\fill}}r@{\extracolsep{0pt}}}
\ifthenelse{\boolean{pdflatex}}
{\vspace*{-2.7cm}\mbox{\!\!\!\includegraphics[width=.14\textwidth]{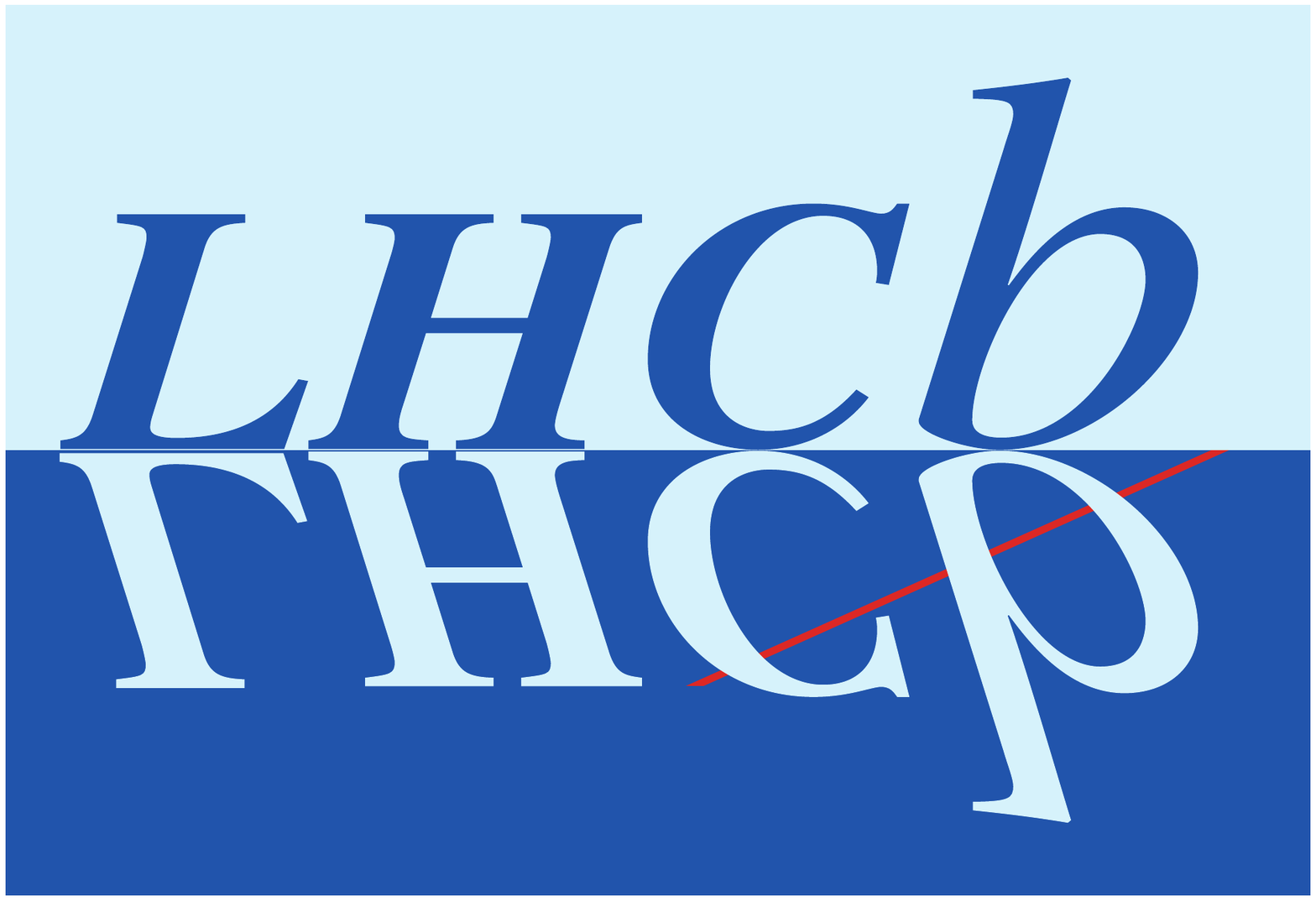}} & &}%
{\vspace*{-1.2cm}\mbox{\!\!\!\includegraphics[width=.12\textwidth]{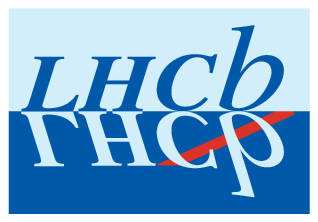}} & &}%
\\
 & & CERN-EP-2017-325 \\  
 & & LHCb-PAPER-2017-039 \\  
 & & \today \\ 
 & & \\
\end{tabular*}

\vspace*{4.0cm}

{\normalfont\bfseries\boldmath\huge
\begin{center}
  \papertitle 
\end{center}
}

\vspace*{2.0cm}

\begin{center}

\paperauthors\footnote{Authors are listed at the end of this paper.}
\end{center}

\vspace{\fill}

\begin{abstract}
  \noindent
A search for the flavor-changing neutral-current decay $\LcToPmm$ is reported using a data set corresponding to an integrated luminosity of $3.0\invfb$ collected by the LHCb collaboration. No significant signal is observed outside of the dimuon mass regions around the $\phi$ and $\omega$ resonances and an upper limit is placed on the branching fraction  of $\BRof\LcTopmumu < 7.7~(9.6)\times 10^{-8}~{\rm at}~90\%~(95\%)$ confidence level. A significant signal is observed in the $\omega$ dimuon mass region for the first time.

\end{abstract}

\vspace*{2.0cm}

\begin{center}
  Published in Phys. Rev. D97 (2018) 091101(R)
\end{center}

\vspace{\fill}

{\footnotesize 
\centerline{\copyright~\papercopyright, licence \href{\paperlicenceurl}{\paperlicence}.}}
\vspace*{2mm}

\end{titlepage}

\newpage
\setcounter{page}{2}
\mbox{~}

\cleardoublepage

\renewcommand{\thefootnote}{\arabic{footnote}}
\setcounter{footnote}{0}



\pagestyle{plain} 
\setcounter{page}{1}
\pagenumbering{arabic}


%

\clearpage

The flavor-changing neutral-current (FCNC) decay  \LcToPmm (inclusion of the charge-conjugate processes is implied throughout) is expected to be heavily suppressed in the Standard Model (SM) by the Glashow-Iliopoulos-Maiani mechanism~\cite{GIM}. The branching fractions for short-distance $c\to u \ellell$ contributions to the transition are expected to be of $\mathcal{O}(10^{-9})$ in the SM, but can be enhanced by effects beyond the SM. However, long-distance contributions proceeding via a tree-level amplitude, with an intermediate meson resonance decaying into a dimuon pair~\cite{FAJFER200393,CHARM_RARE_BURDMAN}, can increase the branching fraction up to $\mathcal{O}(10^{-6})$~\cite{Fajfer:2005ke}. The short-distance and hadronic contributions can be separated by splitting the data set into relevant regions of dimuon mass. The \LcToPmm decay has been previously searched for by the BaBar collaboration~\cite{BABAR} yielding $11.1\pm 5.0 \pm 2.5$ events and an upper limit on the branching fraction of $4.4\times 10^{-5}$ at 90\% CL. 

Similar FCNC transitions for the $\bquark$-quark system ($b\to s \ellell$) exhibit a pattern of consistent deviations from the current SM predictions both in branching fractions~\cite{Aaij:2016flj} and angular observables~\cite{LHCb-PAPER-2015-051}, with the combined significance reaching 4 to 5 standard deviations\mbox{\cite{Capdevila:2017bsm,Altmannshofer:2017fio}}. Processes involving $c \to u \ellell$ transitions are far less explored at both the experimental and theoretical levels, which makes such measurements desirable. Similar analyses of the $D$ system have reported evidence for the long-distance contribution~\cite{LHCb-PAPER-2017-019}, however, the short-distance contributions have not been established~\cite{Aaij:2013sua}.

In this paper we report on the search for the \LcTopmumu decay, using a data set corresponding to an integrated luminosity of $3.0\invfb$ of $pp$ collisions collected in 2011 and 2012 with the LHCb experiment. The branching fraction is measured  with respect to the branching fraction of the decay \LcTopphiNOMU with $\Pphi(1020) \to \mumu$ (here and after denoted as \LcTopphi) decay, which has the benefit of having the same initial and final states and consequently many sources of systematic uncertainty are expected to cancel.

The \lhcb detector~\cite{Alves:2008zz,LHCb-DP-2014-002} is a single-arm forward
spectrometer covering the \mbox{pseudorapidity} range $2<\eta <5$,
designed for the study of particles containing \bquark or \cquark
quarks. The detector includes a high-precision tracking system
consisting of a silicon-strip vertex detector surrounding the $pp$
interaction region~\cite{LHCb-DP-2014-001}, a large-area silicon-strip detector located
upstream of a dipole magnet with a bending power of about
$4{\mathrm{\,Tm}}$, and three stations of silicon-strip detectors and straw
drift tubes~\cite{LHCb-DP-2013-003} placed downstream of the magnet.
The tracking system provides a measurement of momentum, of charged particles with
a relative uncertainty that varies from 0.5\% at low momentum to 1.0\% at 200\gevc.
The minimum distance of a track to a primary vertex (PV), the impact parameter (IP), 
is measured with a resolution of $(15+29/\pt)\mum$,
where \pt is the component of the momentum transverse to the beam, in\,\gevc.
Different types of charged hadrons are distinguished using information
from two ring-imaging Cherenkov detectors~\cite{LHCb-DP-2012-003}. 
Photons, electrons and hadrons are identified by a calorimeter system consisting of
scintillating-pad and preshower detectors, an electromagnetic
calorimeter and a hadronic calorimeter. Muons are identified by a
system composed of alternating layers of iron and multiwire
proportional chambers~\cite{LHCb-DP-2012-002}.
The online event selection is performed by a trigger~\cite{LHCb-DP-2012-004}, 
which consists of a hardware stage, based on information from the calorimeter and muon
systems, followed by a software stage, which applies a full event
reconstruction.

Samples of simulated events are used to understand the properties of the signal and normalization channels.  
The $pp$ collisions are generated using
\pythia~\cite{Sjostrand:2006za,*Sjostrand:2007gs} with a specific \lhcb
configuration~\cite{LHCb-PROC-2010-056}. Decays of hadronic particles are described by \evtgen~\cite{Lange:2001uf}, in which final-state radiation is generated using \photos~\cite{Golonka:2005pn}. The decay of the \Lc baryon to \proton \mumu is simulated with a three-body phase-space model. The interaction of the generated particles with the detector, and its response,
are implemented using the \geant toolkit~\cite{Allison:2006ve, *Agostinelli:2002hh} as described in Ref.~\cite{LHCb-PROC-2011-006}.  The $\Lc$ baryons  are produced in two ways at a hadron collider; as prompt $\Lc$ or in $b$--hadron decays. The simulation contains a mixture of these two production mechanisms, according to the known $\Lc$ and $b$--hadron production cross-sections~\cite{Aaij:2013noa,Aaij:2013mga}.

The simulated samples are used to determine the selection criteria, in particular to train a multivariate classifier that is aimed at distinguishing signal signatures in the background-dominated data set. The simulated samples are also used to calculate the efficiencies of several selection steps.

Candidate events of  \LcToPmm decay are reconstructed by combining a pair of charged tracks identified as muons with one identified as a proton. Candidates that pass the trigger selections are subject to further requirements consisting of kinematic and particle identification criteria and based on the response of a multivariate classifier. Each of the final-state tracks is required to be of good quality, to have $\pt > 300\mevc$ and to be incompatible with originating from any of the PVs in the event. The tracks are also required to form a good-quality secondary vertex with a corresponding flight distance of at least $0.1~\rm mm$ from all of the PVs in the event. The invariant mass of the dimuon system is required to be smaller than $1400\mevcc$.
Three dimuon mass regions are defined:
\begin{itemize}
\item a region around the known $\phi$ mass, $\left[985, 1055\right]\mevcc$, used as a normalization channel;
\item a region around the known $\omega$ mass (the $\omega$ denotes hereafter the $\omega(782)$ meson), $\left[759 ,  805\right]\mevcc$, used to isolate the $\Lcpomegano$ decay;
\item a nonresonant region $(\LcTopmumu)$, with excluded ranges $\pm 40\mevcc$ around the known $\omega$ and $\phi$ masses.
\end{itemize}

After the preselection, the normalization channel is still dominated by combinatorial background, \ie combinations of tracks that do not all originate from a genuine \Lc baryon. 
A boosted decision tree (BDT) is trained to reduce the combinatorial background to manageable level. The BDT is trained using the kinematic and topological variables of the \Lc candidate, related to its flight distance, decay vertex quality, \pt and impact parameter with respect to the primary vertex. In the BDT training, \LcTopmumu simulated events are used as a proxy for the signal and data outside the signal \proton \mumu invariant mass region extending up to $\pm300\mevcc$ around the known $\Lc$ mass is used as a proxy for the background. 

A k-folding technique is used to ensure the training is unbiased~\cite{Blum:1999:BHB:307400.307439}, while keeping the full available data sets for further analysis. A loose BDT cut is applied to reduce the background to the same level as the normalization channel yield. 

\begin{figure}[!htb]
\begin{center}
\includegraphics[width=0.6\textwidth]{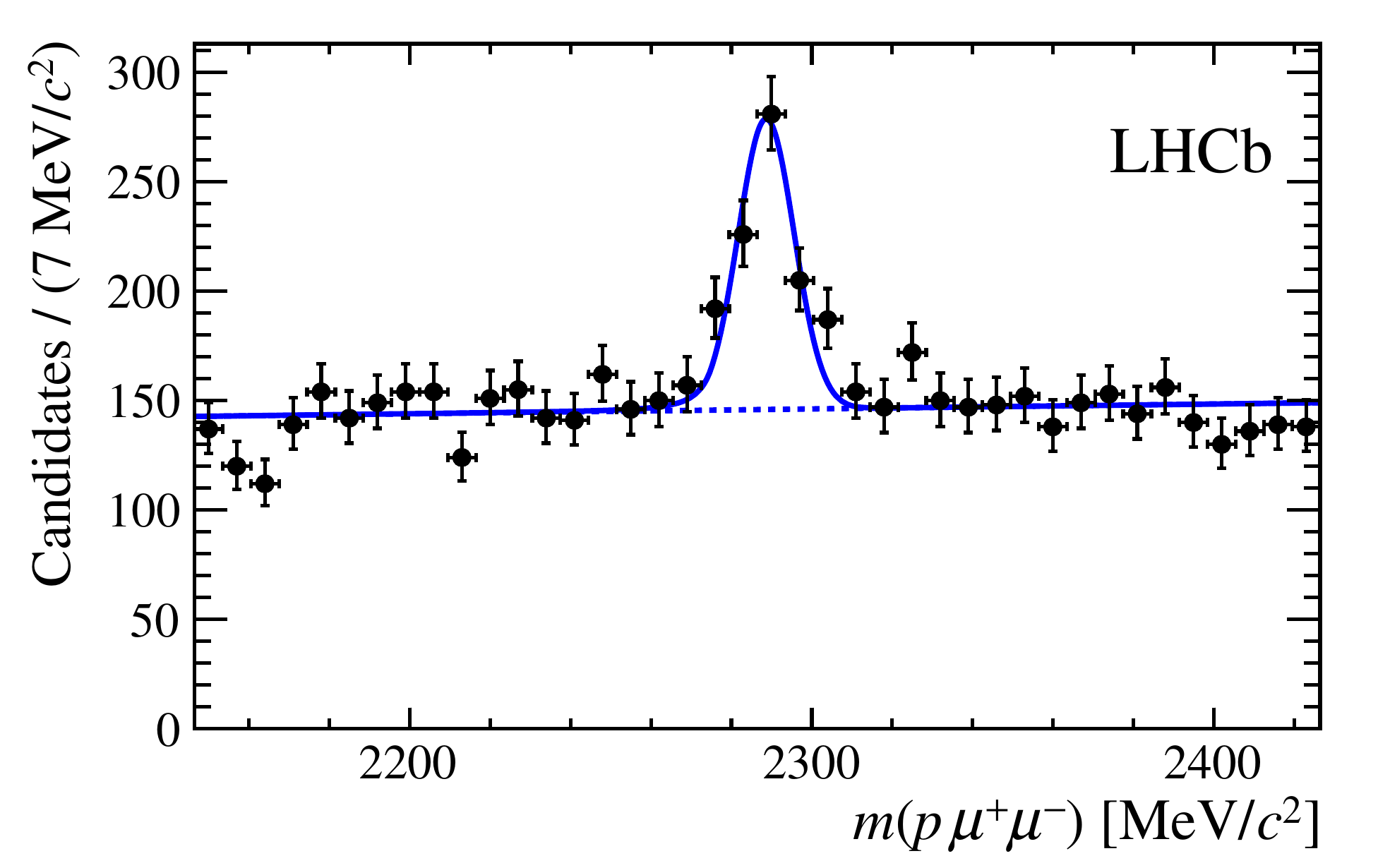}
\end{center}
\caption{Mass distribution of $\LcTopmumu$ candidates in the $\phi$ region after the first BDT requirement. The solid line shows the result of the fit described in the text, while the dashed line indicates the background component.}\label{fig:phi}
\end{figure}

A fit to the $p\mumu$ invariant-mass distribution of \LcTopphi candidates after the loose BDT requirement is shown in Fig.~\ref{fig:phi}. The shape of the \Lc peak is parametrized by a Crystal Ball function~\cite{Skwarnicki:1986xj} with parameters determined from the simulation, while the background is modeled with a first-order polynomial. The yield of the \LcTopphi decay is determined to be $395 \pm 45$ candidates. This sample is used for the final optimization of selection requirements.
It is checked at this stage that the variables used in the signal selection are well described by simulation within the available sample size.

For the final selection a second BDT is trained, which includes additional variables related to \Lc-baryon decay properties and the isolation of the proton and muons in the detector. The final discrimination is performed in three dimensions: the BDT variable and two particle identification (PID) variables, the proton-identification discriminant and the muon-identification discriminant. The optimal set of BDT and PID requirements is determined by finding the best expected upper limit on the branching fraction of the signal relative to the normalization channel using the \CLs method~\cite{CLs} by means of Monte Carlo methods. 

Several sources of background have been considered. An irreducible background due to long-distance contributions originates from
$\Lc \to \proton V(\mumu)$ decays, with intermediate resonances indicated by $V$. The $\rho(770)^0$, $\omega$ and $\eta$ resonances are studied, however, their contribution to the nonresonant region is expected to be negligible, because the $V$ meson mass is well separated from the nonresonant region and/or the $\Lc \to \proton V(\mumu)$ branching fraction is small.
Another background source considered is due to misidentification of final-state particles in hadronic \Dp, \Ds and \Lc decays. The expected contribution from this source has been estimated using large samples of simulated events. Given the tight PID requirements obtained from the optimization, only $2.0 \pm 1.1$ candidates are expected to fall into the \Lc mass window in the nonresonant region.

The ratio of branching fractions is measured using
\begin{equation}                                                                                                                                                                                                                              
\frac{\BRof\LcTopmumu}{  \BRof\Lcpphino\BRof\phimumu} =
\frac{\rm{\epsilon\mathstrut_{norm}}} {\rm {\epsilon\mathstrut_{sig}}}\times\frac{N_{\rm sig}}{N_{\rm norm}},
\label{eq:normalization}
\end{equation}
\noindent where $N_{\rm sig}$ ($N_{\rm norm}$) is the observed yield for the signal (normalization) decay mode. The factors $\rm\epsilon\mathstrut_{sig}$ and $\rm\epsilon\mathstrut_{norm}$ indicate the corresponding total efficiencies for signal and normalization channels, respectively. The efficiencies are determined from the simulation.


In the case of the observation of the decay \LcpV, the ratio of branching fractions is determined by
\begin{equation}
\frac{\BRof\LcpV\BRof{V\to\mumu}}{ \BRof\Lcpphino\BRof\phimumu } = 
\frac{\rm  {\epsilon\mathstrut_{norm}}}{\rm  {\epsilon_{V}}}  \times \frac{N_{V}}{N_{\rm norm}}, 
\label{eq:normalizationomega} 
\end{equation} 
\noindent where $N_{V}$ ($N_{\rm norm}$) is the number of candidates observed for the \LcpV (normalization) decay mode. The factors $\rm\epsilon\mathstrut_{V}$ and $\rm\epsilon\mathstrut_{norm}$ indicate the corresponding total efficiencies for \LcpV and normalization channel, respectively.

As the final states of the signal and normalization channels are identical, many sources of systematic uncertainty cancel in the ratio of the efficiencies. There are three significant sources of systematic uncertainty. The first is related to the finite size of the simulation samples, which limits the precision on the efficiency ratio. The second is linked to residual differences between data and simulation of the BDT distribution. The third is associated to the simulation of PID and is determined from the uncertainty on the PID calibration samples. The values of the contributions are given in Table~\ref{tab:systematic}.

Several other sources of systematic uncertainty were considered: the trigger efficiency, the shapes used in the invariant mass fit for signal and normalization channels, the shape of the combinatorial background, and the fraction of prompt \Lc baryons and \Lc baryons from $b$--hadron decays. All of these, however, are at negligible level when compared to three dominant sources of systematic uncertainty.

\begin{table}[tb]
  \caption{
    Systematic uncertainties on the efficiency ratio used in the determination of the branching fraction in the nonresonant and  $\omega$ regions. }
\begin{center}\begin{tabular}{lcc}
    \hline
    \hline
    Uncertainty source               &   Value [\%]   & Value [\%]    \\ 
                                     &    \LcTopmumu  &  \LcpVmm   \\ 
                                        &    nonresonant  &  $\omega$ region   \\ 
    \hline
    Size of simulation samples             &  4.4    & 10.0  \\
    BDT cut                                    &  4.8    &  ~4.8  \\
    PID cut                                    &  0.7   &  ~0.7 \\\hline
	Total                                   &  6.5   &  11.1 \\
    \hline
    \hline
  \end{tabular}\end{center}
\label{tab:systematic}
\end{table}

\begin{figure}[!htb]
\begin{center}
\includegraphics[width=0.6\textwidth]{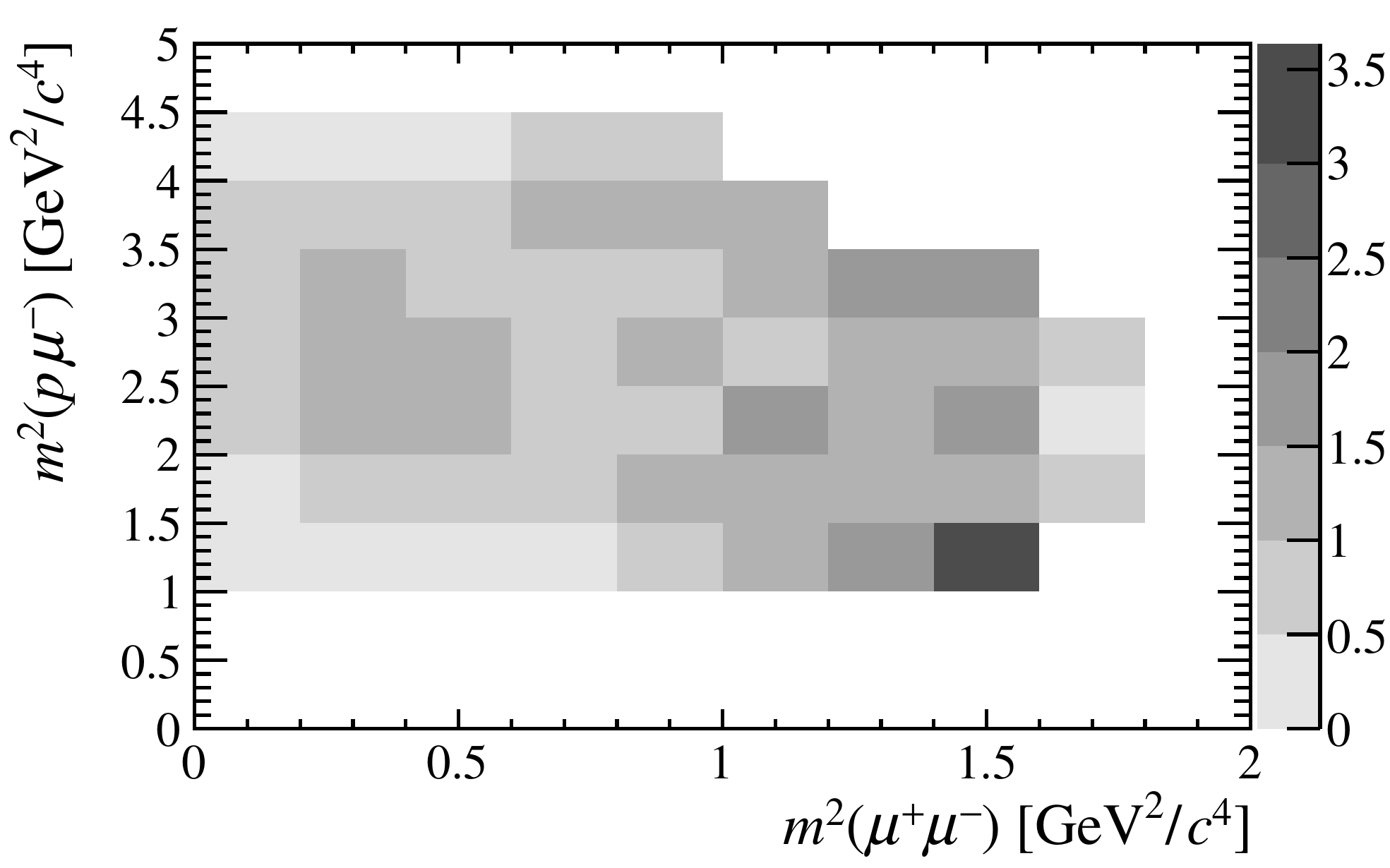}
\end{center}
\caption{The efficiency weights for \LcTopmumu as a function of the dimuon invariant mass squared $m^{2}(\mumu)$ and by the invariant mass of the proton and the negatively charged muon squared $m^{2}(\proton\mu^-)$. The weights are normalized to the average efficiency.}
\label{fig:dalitz_weights}

\end{figure}

The simulated \LcTopmumu decays have been generated according to a phase-space model for the decay products. As the exact physics model for the decay is not known, no systematic uncertainty is assigned. Instead, the weights needed to recast the result in terms of any physics model are provided in~Fig.~\ref{fig:dalitz_weights}. The weights are described by a function of the dimuon invariant mass squared $m^{2}(\mumu)$ and the invariant mass of the proton and the negatively charged muon squared $m^{2}(\proton\mu^-)$. The weights are normalized to the average efficiency. 

\begin{figure}[!htb]
\begin{center}
\includegraphics[width=0.6\textwidth]{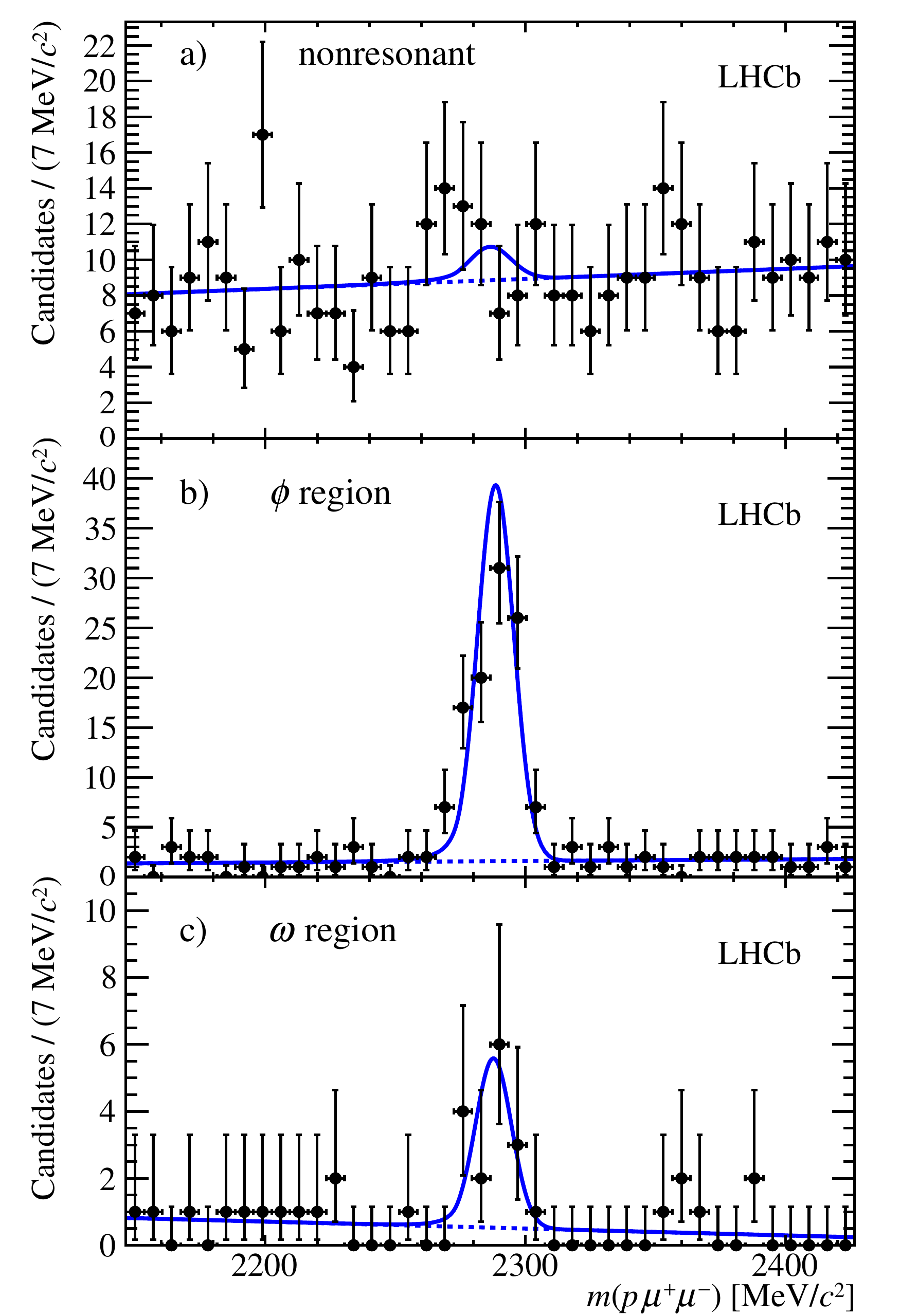}
\end{center}
\caption{Mass distribution for selected  \proton \mumu candidates in the three regions of the dimuon invariant mass: a) nonresonant region, b) $\phi$ region, c) $\omega$ region. The solid lines show the results of the fit as described in the text. The dashed lines indicate the background component.} 
\label{fig:signal}

\end{figure}

The distributions of the \proton \mumu invariant mass for the \LcTopmumu candidates after final selections in the three dimuon mass ranges are presented in Fig.~\ref{fig:signal}. The \Lc peak is parametrized by a Crystal Ball~\cite{Skwarnicki:1986xj} function with parameters determined from the simulation and the background is described by a first-order polynomial. The fits are used to determine the signal yields.
No significant signal is observed in the nonresonant region (Fig.~\ref{fig:signal}a). The yield for the normalization channel is determined to be $96 \pm 11$ candidates (Fig.~\ref{fig:signal}b). An accumulation of $13.2\pm4.3$ candidates at the \Lc mass is observed in the $\omega$ region (Fig.~\ref{fig:signal}c). The statistical significance of the excess is determined to be $5.0\sigma$ using Wilks' theorem~\cite{wilks1938}. 

\begin{figure}[tbh]
  \begin{center}
    \includegraphics[width=0.6\textwidth]{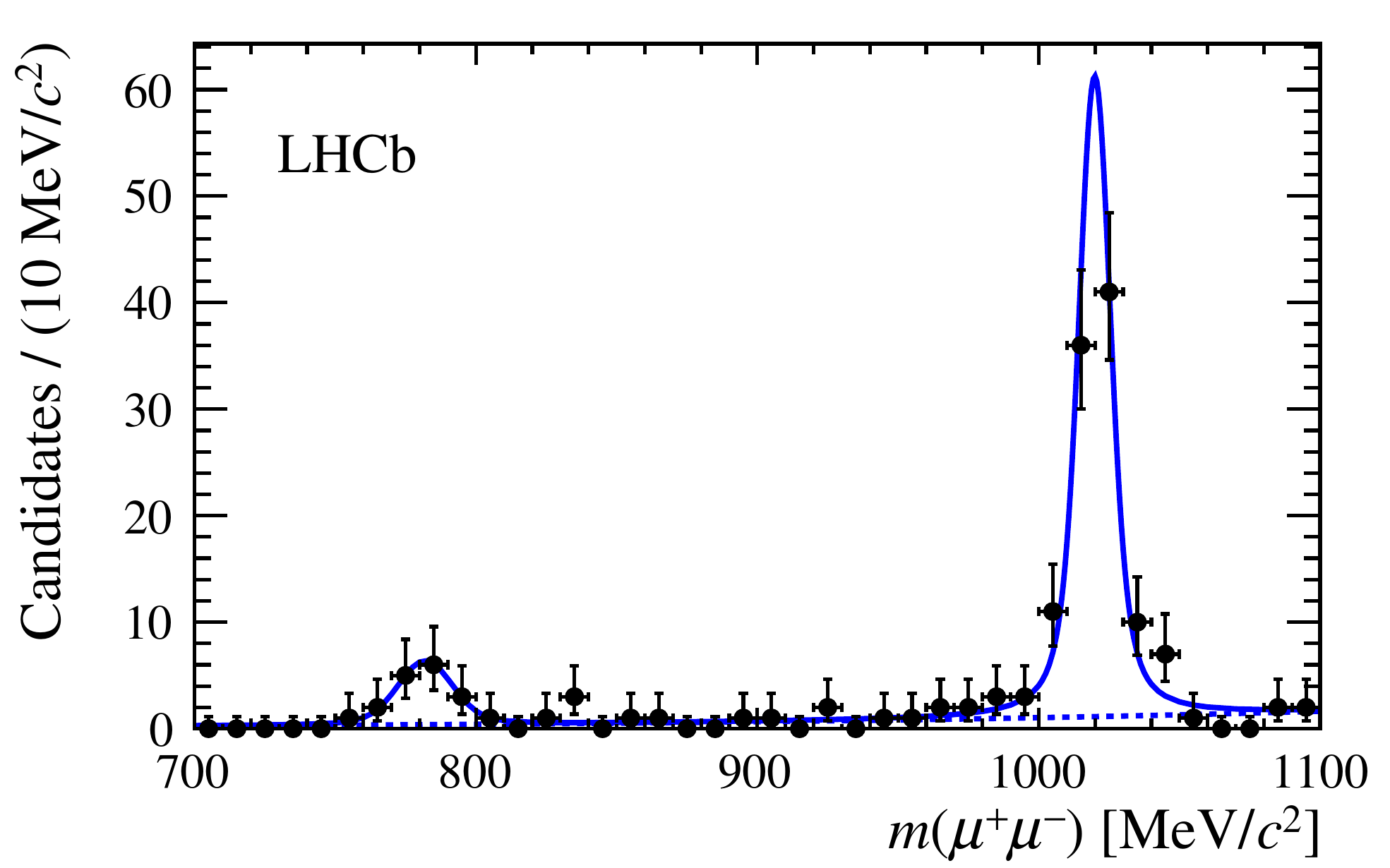}
  \end{center}
  \caption{Invariant mass distribution  $m({\mumu})$ for $\LcTopmumu$ candidates with mass $\pm 25\mevcc$ around the \Lc mass.  The solid line shows the result of the fit, while the dashed line indicates the background component.} 
  \label{fig:massmumu}
\end{figure}

The distribution of the dimuon invariant mass of the \Lc candidates 
is shown in Fig.~\ref{fig:massmumu}. 
An excess is seen at the known $\omega$ and $\phi$  resonance masses. The data is well described by a simple model including these resonances and a background component. The $\omega$ and $\phi$ peaks are parametrized as Breit-Wigner functions of relevant decay width~\cite{PDG2016} convolved with a Gaussian function to take into account the experimental resolution. The addition of a component for the $\rho(770)^0$ resonance (and its interference with the $\omega$ meson) does not improve the fit quality. It is therefore assumed that the observed candidates in the $\omega$ region are dominated by decays via the $\omega$ resonance.

\begin{figure}
  \begin{center}
    \includegraphics[width=0.6\textwidth]{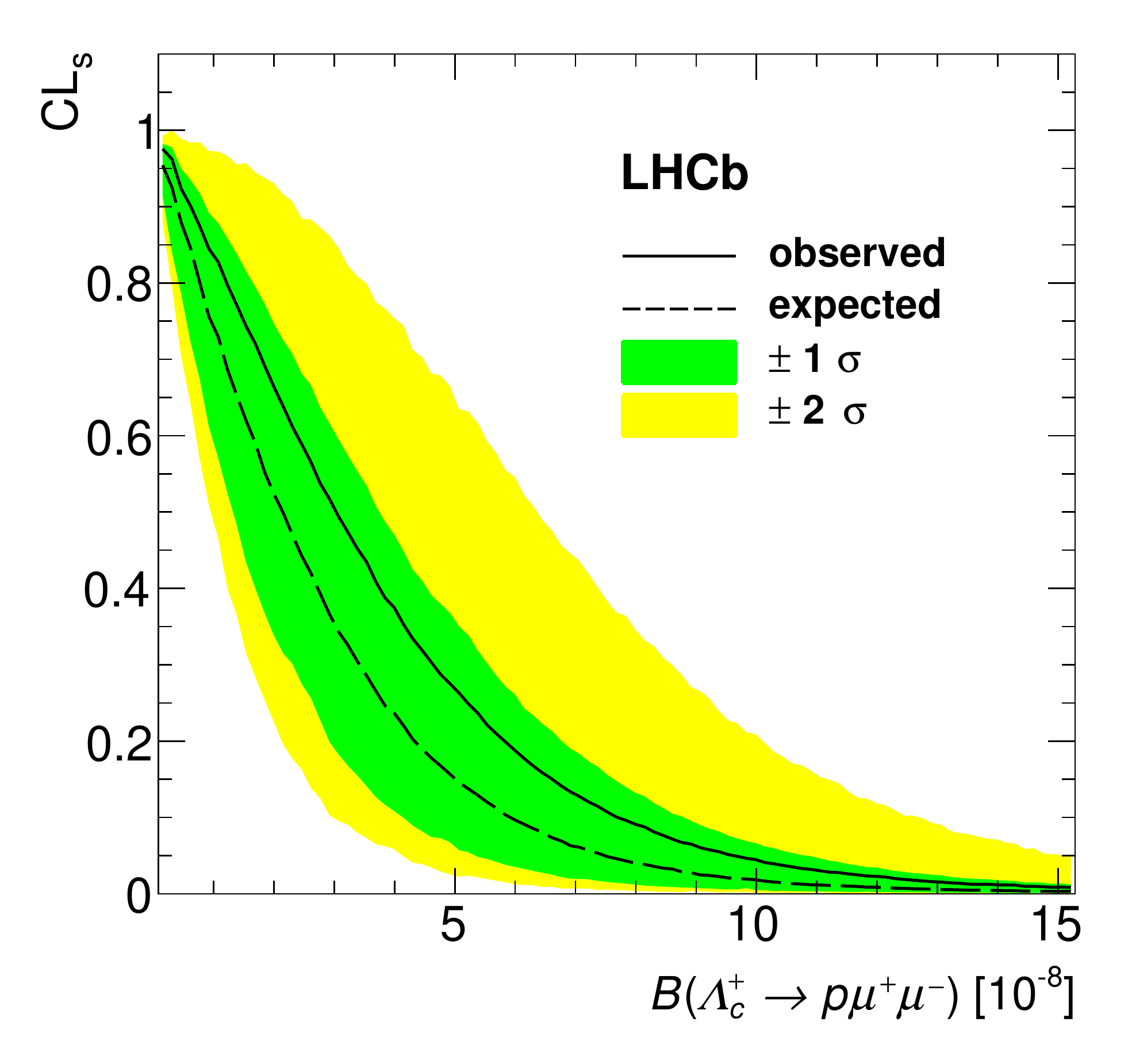}
  \end{center}
  \caption{The \CLs value as a function of the $\BRof\LcTopmumu$ branching fraction. The median expected value
  of an ensemble (assuming no signal component) is shown by the dashed line, with the $\pm 1\sigma$ and $\pm 2\sigma$ regions shaded. The observed distribution is shown by the solid line.}
  \label{fig:clsbf}
\end{figure}

As no evidence for nonresonant $\LcTopmumu$ decays is found, an upper limit on the branching fractions is determined using the \CLs method. The systematic uncertainties are included in the construction of \CLs. The following upper limits are obtained at different confidence levels (CL)
\begin{align*}                                                                                                                                             \frac{ \BRof\LcTopmumu }{  \BRof\Lcpphino\BRof\phimumu  }  < 0.24~(0.28)~~~~{\rm at}~~90\%~(95\%)~{\rm CL}
\end{align*}                                    
The corresponding distribution of \CLs is shown in Fig.~\ref{fig:clsbf}.
Using the values of the branching fractions for \Lcpphino and \phimumu decays from Ref.~\cite{PDG2016} and including their uncertainties in the \CLs construction, an upper limit on the branching fraction is determined to be 
\begin{align*} 
\BRof\LcTopmumu < 7.7~(9.6) \times 10^{-8}~~~~{\rm at}~~90\%~(95\%)~{\rm CL}.
 \end{align*} 

Under the above-mentioned assumption of the \Lcpomegano dominance in the $\omega$ region, the relative branching fraction with respect to the normalization channel is determined according to Eq.~\ref{eq:normalizationomega}

\begin{align*}
\frac{\BRof\Lcpomegano\BRof\omegamumu }{ \BRof\Lcpphino\BRof\phimumu } = 0.23 \pm 0.08~({\rm stat}) \pm 0.03~({\rm syst}).~~~
\end{align*}
Using the relevant branching fractions from Ref.~\cite{PDG2016}, the branching fraction of \Lcpomegano is determined to be
\begin{align*}                                                                                                                                                
\BRof\LcTopomegaNM  = (9.4 \pm 3.2~({\rm stat}) \pm 1.0~({\rm syst}) \pm 2.0~({\rm ext}) )\times 10^{-4},~~~                                                                  
\end{align*}                                                                                                                                                  
where the first uncertainty is statistical, the second corresponds to the above-mentioned systematic effects and the third is due to the limited knowledge of the relevant branching fractions. Assuming lepton universality, the branching fraction \BRof{\omega \to e^{+}e^{-}} is used instead of $\BRof\omegamumu$. 

In summary, a search for the \LcTopmumu decay is reported, using $pp$ data collected with the LHCb
experiment. The analysis is performed in three regions of dimuon mass: $\phi$, $\omega$ and nonresonant. The upper limit on the nonresonant mode is improved by two orders of magnitude with respect to the previous measurement~\cite{BABAR}. For the first time the signal is seen in the $\omega$ region with a statistical significance of 5 standard deviations.


\section*{Acknowledgements}
%
%
\noindent We express our gratitude to our colleagues in the CERN
accelerator departments for the excellent performance of the LHC. We
thank the technical and administrative staff at the LHCb
institutes. We acknowledge support from CERN and from the national
agencies: CAPES, CNPq, FAPERJ and FINEP (Brazil); MOST and NSFC
(China); CNRS/IN2P3 (France); BMBF, DFG and MPG (Germany); INFN
(Italy); NWO (The Netherlands); MNiSW and NCN (Poland); MEN/IFA
(Romania); MinES and FASO (Russia); MinECo (Spain); SNSF and SER
(Switzerland); NASU (Ukraine); STFC (United Kingdom); NSF (USA).  We
acknowledge the computing resources that are provided by CERN, IN2P3
(France), KIT and DESY (Germany), INFN (Italy), SURF (The
Netherlands), PIC (Spain), GridPP (United Kingdom), RRCKI and Yandex
LLC (Russia), CSCS (Switzerland), IFIN-HH (Romania), CBPF (Brazil),
PL-GRID (Poland) and OSC (USA). We are indebted to the communities
behind the multiple open-source software packages on which we depend.
Individual groups or members have received support from AvH Foundation
(Germany), EPLANET, Marie Sk\l{}odowska-Curie Actions and ERC
(European Union), ANR, Labex P2IO and OCEVU, and R\'{e}gion
Auvergne-Rh\^{o}ne-Alpes (France), RFBR, RSF and Yandex LLC (Russia),
GVA, XuntaGal and GENCAT (Spain), Herchel Smith Fund, the Royal
Society, the English-Speaking Union and the Leverhulme Trust (United
Kingdom).



\addcontentsline{toc}{section}{References}
\setboolean{inbibliography}{true}
\bibliographystyle{LHCb}
 \ifx\mcitethebibliography\mciteundefinedmacro
\PackageError{LHCb.bst}{mciteplus.sty has not been loaded}
{This bibstyle requires the use of the mciteplus package.}\fi
\providecommand{\href}[2]{#2}

\newpage


 
\newpage
\centerline{\large\bf LHCb collaboration}
\begin{flushleft}
\small
R.~Aaij$^{40}$,
B.~Adeva$^{39}$,
M.~Adinolfi$^{48}$,
Z.~Ajaltouni$^{5}$,
S.~Akar$^{59}$,
J.~Albrecht$^{10}$,
F.~Alessio$^{40}$,
M.~Alexander$^{53}$,
A.~Alfonso~Albero$^{38}$,
S.~Ali$^{43}$
% ...

{\footnotesize \it
% ...
$ ^{5}$Clermont Universit{\'e}, Universit{\'e} Blaise Pascal, CNRS/IN2P3, LPC, Clermont-Ferrand, France\\
% ...
$ ^{10}$Fakult{\"a}t Physik, Technische Universit{\"a}t Dortmund, Dortmund, Germany\\
% ...
$ ^{38}$ICCUB, Universitat de Barcelona, Barcelona, Spain\\
$ ^{39}$Instituto Galego de F{\'\i}sica de Altas Enerx{\'\i}as (IGFAE), Universidade de Santiago de Compostela, Santiago de Compostela, Spain\\
$ ^{40}$European Organization for Nuclear Research (CERN), Geneva, Switzerland\\
% ...
$ ^{43}$Nikhef National Institute for Subatomic Physics, Amsterdam, The Netherlands\\
% ...
$ ^{48}$H.H. Wills Physics Laboratory, University of Bristol, Bristol, United Kingdom\\
% ...
$ ^{53}$School of Physics and Astronomy, University of Glasgow, Glasgow, United Kingdom\\
% ...
$ ^{59}$University of Cincinnati, Cincinnati, OH, United States\\
% ...
}
\end{flushleft}
\clearpage

\end{document}